\documentclass[a4paper]{PoS}
\usepackage[utf8]{inputenc}
\usepackage{amsmath}
\usepackage{xspace}
\usepackage{bm}
\usepackage{cite}
\usepackage{microtype}

\newcommand{\fire}{\textsc{FIRE}\xspace}
\newcommand{\mathematica}{\textsc{Mathematica}\xspace}
\newcommand{\form}{\textsc{FORM}\xspace}
\newcommand{\cpp}{\textsc{C++}\xspace}
\newcommand{\qgraf}{\textsc{QGraf}\xspace}
\newcommand{\litered}{\textsc{LiteRed}\xspace}
\newcommand{\fuchsia}{\textsc{Fuchsia}\xspace}
\newcommand{\hpl}{\textsc{HPL}\xspace}
\newcommand*{\cf}{cf.\ }
\newcommand*{\eg}{e.\,g.\ }
\newcommand*{\Eq}{Eq.\,}
\newcommand*{\Fig}{Fig.\,}

\allowdisplaybreaks

\title{Analytic calculation of Energy-Energy Correlation in $e^+ e^-$ annihilation at NLO}
\ShortTitle{Energy-Energy-Correlation at NLO}

\author{Lance~J.~Dixon\\
        SLAC National Accelerator Laboratory, Stanford
        University, Stanford, CA 94039, USA\\
        E-mail: \email{lance@slac.stanford.edu}}

\author{Ming-xing~Luo\\
        Zhejiang Institute of Modern Physics, Department of
        Physics, Zhejiang University, Hangzhou, 310027, China\\
        E-mail: \email{mingxingluo@zju.edu.cn}}

\author{\speaker{Vladyslav Shtabovenko}\\
        Zhejiang Institute of Modern Physics, Department of
        Physics, Zhejiang University, Hangzhou, 310027, China\\
        E-mail: \email{vshtabov@zju.edu.cn}}

\author{Tong-Zhi~Yang\\
        Zhejiang Institute of Modern Physics, Department of
        Physics, Zhejiang University, Hangzhou, 310027, China\\
        E-mail: \email{yangtz@zju.edu.cn}}

\author{Hua~Xing~Zhu\\
        Zhejiang Institute of Modern Physics, Department of
        Physics, Zhejiang University, Hangzhou, 310027, China\\
        E-mail: \email{zhuhx@zju.edu.cn}}

\abstract{We present the first fully analytic calculation of the Quantum Chromodynamics (QCD) event shape observable Energy-Energy Correlation in electron-positron annihilation at Next-To-Leading Order (NLO). This result sheds light on the analytic structure of the event shape observables beyond Leading Order (LO) and serves as a motivation to employ our methods in the investigation of other event shape observables that so far have not been calculated analytically.}

\FullConference{Loops and Legs in Quantum Field Theory (LL2018)\\
                29 April 2018 - 04 May 2018\\
                St. Goar, Germany}

\begin{document}

\section{Introduction}

Plans for future electron-positron colliders are currently being put forward in Europe (FCC-ee\cite{Gomez-Ceballos:2013zzn}, CLIC\cite{Aicheler:2012bya}), China (CEPC \cite{CEPC-SPPCStudyGroup:2015csa}) and Japan (ILC\cite{Behnke:2013xla} ). These machines would provide us with high-precision measurements of various Standard Model (SM) observables, which could help not only to scrutinize the properties of the recently discovered scalar boson, but also to improve our understanding of the strong force.

The latter can be approached by studying QCD event shape observables such as Thrust \cite{Brandt:1964sa, Farhi:1977sg}, $C$-parameter \cite{Parisi:1978eg, Donoghue:1979vi,Ellis:1980wv}, Jet Broadening \cite{Rakow:1981qn, Ellis:1986ig,Catani:1992jc}, Jet Masses \cite{Clavelli:1979md} or Energy-Energy Correlation (EEC) \cite{Basham:1977iq}. The event shape observables represent a special set of infrared (IR) safe quantities that were designed to be equally well accessible through experimental measurements and calculations in perturbative QCD (pQCD). In the following, we will focus on the EEC, which can be defined by considering electron-positron annihilation into two hadrons plus anything $e^+ e^- \to a + b + X$ and measuring the energies of the hadrons $a$ and $b$ with two calorimeters separated by the angle $\chi$. EEC corresponds to the differential angular distribution  of the energy flowing through these calorimeters and is defined as
\begin{equation}
  \frac{d \Sigma  (\chi)}{d \cos\chi} =
\sum_{a,b} \int \, \frac{E_a E_b}{Q^2} \, \delta( \cos\theta_{ab} -
  \cos\chi) \,  d \sigma_{a+b+X}, \quad  \cos \theta_{ab} = \hat{\bm{p}}_a \cdot \hat{\bm{p}}_b, \label{eq:eecdef}
\end{equation}
with $Q^2$ being the total CM energy squared, while $\hat{\bm{p}}_a$ and $\hat{\bm{p}}_b$ denote the directions of the hadron 3-momenta.   In pQCD we can use the corresponding momentum sum rule to calculate this quantity by replacing the hadrons with partons, where the leading contribution arises from the process $e^+ e^- \to q \bar{q} \, g$. Notice that EEC also receives nonperturbative corrections (\cf \cite{Dokshitzer:1999sh} for more details), but those are beyond the scope of this work. 

The analytic LO result for this observable was presented in the original publication \cite{Basham:1977iq} and is given by
\begin{equation}
\frac{1}{\sigma_{\textrm{tot}}} \frac{d \Sigma (\chi)}{d\cos\chi}  =  \frac{\alpha_s(\mu)}{2 \pi} \, C_F  \frac{3 - 2 z}{4 (1 - z) z^5} \Bigl[ 3 z (2 - 3 z) + 2 (2 z^2 - 6 z + 3) \log(1-z) \Bigr] + \mathcal{O}(\alpha_s^2),
\end{equation}
where we introduced $z \equiv (1- \cos \chi)/2$  and $\sigma_{\textrm{tot}}$ is the Born cross section for $e^+ e^- \to q  \bar{q}$. Subsequent NLO calculations were carried out using numerical methods 
\cite{Schneider:1983iu,Falck:1988gb,Glover:1994vz,Kramer:1996qr,Ali:1982ub,Barreiro:1986si,Richards:1982te, Richards:1983sr,Kunszt:1989km,Catani:1996jh,Catani:1996vz, Clay:1995sd}. The publicly available code \textsc{Event2} provides a stable implementation of the dipole subtraction method \cite{Catani:1996jh,Catani:1996vz} and can be used to obtain reliable NLO results. More recently, the NNLO results were derived in the CoLoRfulNNLO subtraction scheme \cite{DelDuca:2016csb,Tulipant:2017ybb}. An important milestone on the way to obtain analytic results beyond LO was the NLO calculation of EEC in the $\mathcal{N}=4$ Supersymmetric Yang-Mills (SYM) theory \cite{Belitsky:2013ofa}.

In this proceeding we briefly report on the first fully analytic result for EEC at NLO. We explain how this result was obtained by combining publicly available tools for automatic calculations in a clever way and discuss the challenges that we faced during this process.  Finally, we present the final result, which exhibits a remarkably simple structure and contains only classical polylogarithms up to weight 3. For details, we refer to our original publication \cite{Dixon:2018qgp}.

\section{Method}
\begin{figure}[ht]
\centering
\includegraphics[width=4cm,clip]{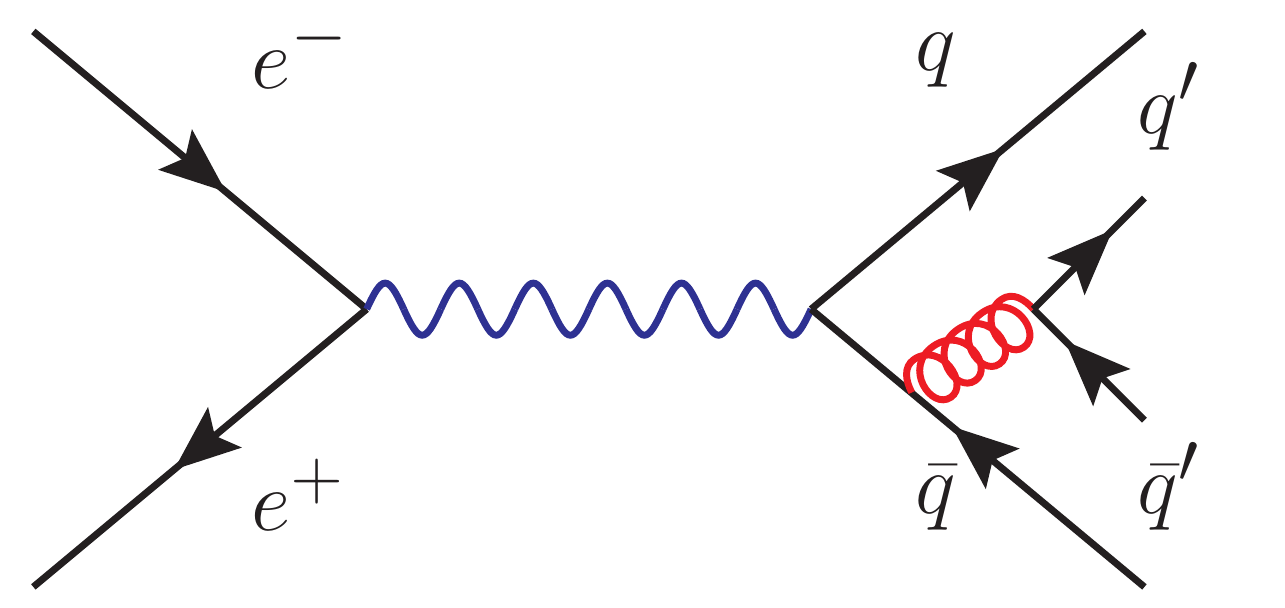}
\includegraphics[width=4cm,clip]{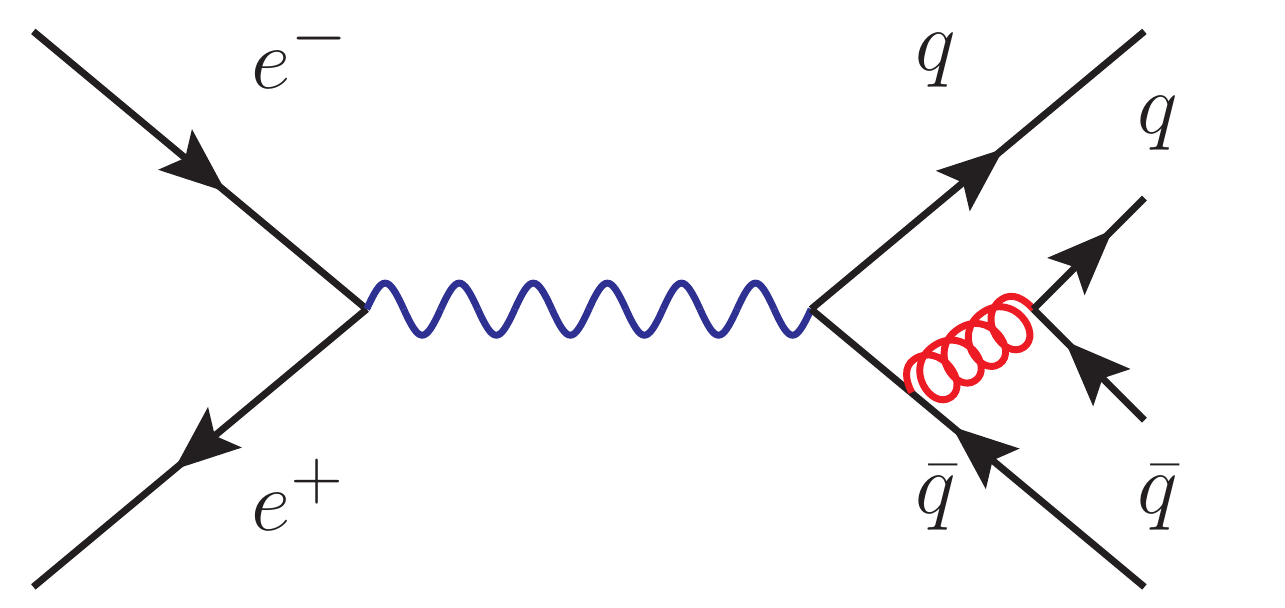}
\includegraphics[width=4cm,clip]{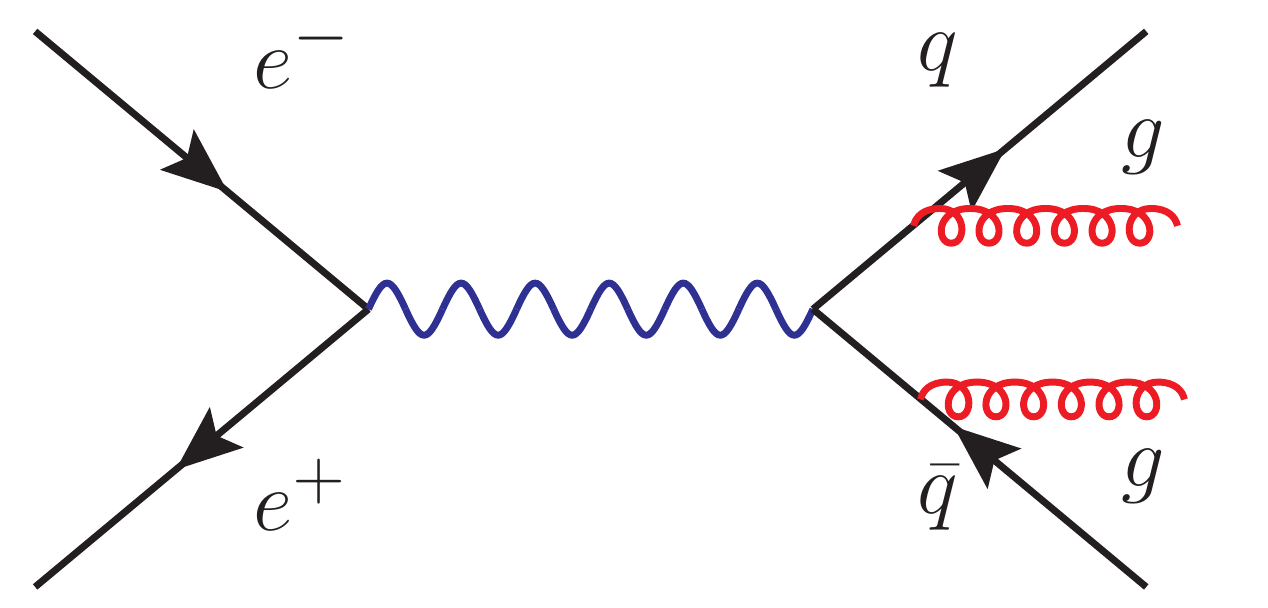}
\caption{Representative QCD diagrams for the three classes of real corrections in EEC.
}
\label{fig:qcd-real}
\end{figure}

The LO contribution\footnote{In our calculation we take all quarks to be massless and neglect the small  \cite{Hagiwara:1990dx} contribution from the $Z$-boson pole.} to the EEC is given by just two tree-level Feynman diagrams,
that describe the electron-positron annihilation into a quark-antiquark pair and a gluon. At NLO one has to consider the virtual correction to $e^+ e^- \to q (p_1) + \bar{q} (p_2) + g (p_3)$ and three classes of real radiative corrections (\cf \Fig\ref{fig:qcd-real}).

The generation of the corresponding Feynman amplitudes and the calculation of the matrix elements squared $\mathcal{M}^2$ can be carried out straightforwardly using \qgraf \cite{Nogueira:1991ex} and \form \cite{Vermaseren:2000nd} together with the \textsc{Color} \cite{vanRitbergen:1998pn} package. The main challenge in obtaining the analytic results for EEC is the evaluation of the phase-space integral in \Eq\ref{eq:eecdef}, where for each combination of the measured particles (\eg $q \bar{q}$ or $q g$), $\mathcal{M}^2$ is multiplied by the corresponding measurement function
\begin{equation}
\left (\tfrac{(Q \cdot p_a) (Q \cdot p_b)}{Q^2} \right)^2 \, \delta \left ( (1 -  \cos\chi) (Q \cdot p_a) (Q \cdot p_b) - Q^2 (p_a \cdot p_b)  \right).
\end{equation}
In the case of the LO and the virtual NLO contributions, the phase-space integrals are finite and can be calculated directly. This method is, however, not applicable to the real radiative corrections, where the phase-space integration leads to unresolved soft and collinear IR divergences.

We choose to approach this difficulty by using the methods of IBP reduction \cite{Chetyrkin:1981qh} and reverse unitarity
\cite{Anastasiou:2002yz,Anastasiou:2003yy} so that we can first rewrite our complicated phase-space integrals as loop integrals and then reduce them to a small set of master integrals. At first sight, this does not make our task any simpler, as the cut propagator obtained from the measurement function turns out to be quadratic in the scalar products involving loop momenta. Since publicly available IBP reduction tools are designed to deal with the (cut) propagators of the type $1/(p^2 \pm m^2)$ or $1/(p \cdot q \pm m^2)$, there seems to be no easy way to apply IBP reduction to integrals with such nonlinear propagators as ${1}/({(1-\cos \chi)(p_a \cdot Q)(p_b \cdot Q) - Q^2 (p_a \cdot p_b))} \bigl |_{\textrm{cut}}$. At this point it is important to realize that this is merely a technical but not a fundamental obstacle. Once we  identify all the unique integral topologies\footnote{The topology identification is carried out using an in-house code.}, nothing prevents us from deriving IBP equations for each of these topologies and then solving them in the usual way. Although this does not work out-of-the box with the existing software tools, it turns out
that a combination of \litered \cite{Lee:2012cn,Lee:2013mka} and \fire \cite{Smirnov:2008iw,Smirnov:2014hma} supplemented with a minimal amount of extra \mathematica code can be used to achieve the desired result with very little effort.

When employing \litered we define each unique topology using the standard \texttt{NewBasis} function, but in such
a way that the nonlinear propagator is omitted. The propagator is then added by  modifying the definition of the \texttt{Ds} function for each topology. This way one can directly derive the corresponding IBP equations via \texttt{GenerateIBP}. The so-obtained
set of equations is, however, still incomplete, and has to supplemented by the additional relation
\begin{equation}
\int \left [ \left ( (1-\cos \chi)(p_a \cdot Q)(p_b \cdot Q) - Q^2 (p_a \cdot p_b) \right ) J_{ab}(n_1,\ldots,n_{10}) - J_{ab}(n_1,\ldots,n_{10}-1)  \right ] = 0,
\end{equation}
where $n_{10}$ is the integer power of the nonlinear propagator in the integrand $J_{ab}$. For each topology this yields a solvable system of the IBP equations that, together with the list of the occurring loop integrals, can be imported into a suitable tool for IBP reduction. Now we only need to find a package that can solve IBP equations and perform the IBP reduction without any reference to the explicit form of the involved propagators, so that the nonlinear propagator will not cause any troubles. Fortunately, this can be accomplished by using \fire. While one would normally generate \fire scripts by specifying all the internal and external momenta and the propagators, 
it is also possible to bypass this step and supply only the set of the corresponding IBP equations. The relevant variable is called \texttt{startinglist} and is explicitly mentioned in the manual of the package \cite{Smirnov:2014hma}.  This allows us to perform the reduction of our highly nonstandard phase-space integrals in a fast and convenient way. Using the \cpp version of \fire on a server equipped with Intel Xeon E5-2696 (18 cores) and $128\textrm{ GB}$ RAM, we can reduce all the 12375 3-loop integrals from the real corrections in just 3 days.

Alternatively, one could also linearize the propagator that stems from the measurement function by introducing
an auxiliary variable $x$
\begin{align}
\delta \left ((1-\cos \chi)(p_a \cdot Q)(p_b \cdot Q) - Q^2 (p_a \cdot p_b) \right ) & =  \int_0^1 d x \, \delta(x- (p_a \cdot Q)) \nonumber \\ 
& \times \delta (x (1-\cos \chi)(p_b \cdot Q) - Q^2 (p_a \cdot p_b) ),
\end{align}
which yields master integrals that depend on two variables $x$ and $z$. This method is used in the analytic calculation of EEC at NLO described in \cite{Gituliar:2017umx,Gituliar:2017vmg}, but as of now the final result obtained with this technique is not yet available.

Our approach leaves us with only 40 master integrals that are, however, too complicated to be evaluated directly. To calculate each of these integrals analytically we turn to the method of differential equations \cite{Kotikov:1991pm,Kotikov:1990kg,Kotikov:1991hm,Bern:1993kr,Remiddi:1997ny,Gehrmann:1999as}. Here we employ the publicly available package \fuchsia \cite{Gituliar:2017vzm} that can automatically turn systems of differential equations into a canonical form \cite{Henn:2013pwa} using Lee's algorithm \cite{Lee:2014ioa}. 
While the package can easily obtain the canonical form for many of our systems of equations, in some cases it fails to find a suitable transformation matrix. The reason for this is that such systems require a nonlinear transformation of variables, which is not covered by the original version of Lee's algorithm. In principle, one can often guess such transformations by looking at the structure of the differential equations and the kinematical constraints of the process under consideration. In our case this turns out to be unnecessary, since we can directly benefit from the exact result for EEC at NLO in $\mathcal{N}=4$ SYM obtained in \cite{Belitsky:2013ofa}.  The presence of letters involving square roots of $z$ and $1-z$ in the  $\mathcal{N}=4$ result already provides a strong hint for the form of the required nonlinear transformations. After some trial and error we identify the two transformations $z \to \sqrt{z}$ and $z \to i \sqrt{z}/\sqrt{1-z}$, which are sufficient to turn all remaining systems of equations into the canonical form using \fuchsia and to solve them in terms of harmonic polylogarithms \cite{Remiddi:1999ew} to any desired order in $\varepsilon$.

The last hurdle to be overcome on the way towards the fully analytic NLO result is the determination of the integration constants. Since the EEC calculated in the fixed-order perturbation theory diverges for $z \to 0$ and $z \to 1$, it is not easy to find suitable boundary conditions that can be used to fix those constants. While we obviously cannot demand the regularity of the master integrals in the collinear limit, it is still possible to use power counting to predict the power of the leading contribution in the expansion around $z=0$. Such expansions of harmonic polylogarithms can be conveniently done using the \hpl package \cite{Maitre:2005uu,Maitre:2007kp} and allow us to determine a large number of the integration constants. We also consider the limit $z \to \infty$, demanding that particular master integrals should have no support for such unphysical values of $z$ and therefore vanish.  Furthermore, the master integrals can be multiplied with
a suitable weight function $z^m (1-z)^n$ with $m,n \in \mathbb{Z}$
and integrated over $z$, which allows us (after performing an additional IBP reduction) to match them to linear combinations of the master integrals of the inclusive 4-particle phase-space \cite{Gehrmann-DeRidder:2003pne}. One should note that this particular boundary condition is similar to the method employed in \cite{Gituliar:2017umx,Gituliar:2017vmg}, although in our case the master integrals do not have an additional dependence on the auxiliary variable $x$, which makes the integration simpler. Finally, we can substitute the preliminary results for the master integrals (with some integration constants still undetermined) into the full result for the real corrections and demand that $1/z$ is the strongest singularity that may appear in the collinear limit. The combination of the above boundary conditions fixes all the relevant constants and allows us to arrive at the final analytic result for EEC at NLO. The first nontrivial test satisfied by this result is the expected complete cancellation of all soft and collinear singularities, which leaves us with a manifestly IR finite expression.

\section{Analytic results}
Our final result for EEC at NLO is given by
\begin{align}
 \frac{1}{\sigma_{\textrm{tot}}} \frac{d\Sigma(\chi)}{d\cos\chi} = &
 \frac{\alpha_s(\mu)}{2\pi} A(z) + \left(\frac{\alpha_s(\mu)}{2 \pi}\right)^2   \left(\beta_0  \log \frac{\mu}{Q} A(z) + B(z) \right)  + \mathcal{O} (\alpha_s^3),
\end{align}
with $\beta_0 = 11C_A/3 -  4 N_f T_f/3$, where the NLO coefficient $B(z)$ admits the following color decomposition
\begin{align}
  B(z) = C_F^2 B_{\text{lc}}(z)  + C_F (C_A - 2 C_F) B_{\text{nlc}}(z)
    + C_F N_f T_f B_{N_f}(z).
\end{align}
All the color coefficients can be written in the basis of pure functions $g_i^{(n)}$ of uniform transcendental weight $n \leq 3$ defined as

{\footnotesize
\begin{align}
  g_1^{(1)} & = \log (1-z)\,, \qquad g_2^{(1)} =  \log (z)\,, \qquad
  g_1^{(2)} = 2 (\text{Li}_2(z)+\zeta_2)+\log ^2(1-z) \,,  \\
  g_2^{(2)} &= \text{Li}_2(1-z)-\text{Li}_2(z) \,, \quad
  g_3^{(2)} = - 2 \, \text{Li}_2\left(-\sqrt{z}\right)
  + 2 \, \text{Li}_2\left(\sqrt{z}\right)
  + \log\left(\frac{1-\sqrt{z}}{1+\sqrt{z}}\right) \log (z) \,,\qquad
  g_4^{(2)} = \zeta_2 \,, \\
  g_1^{(3)} & = -6 \left( \text{Li}_3\left(-\frac{z}{1-z}\right)-\zeta_3 \right)
- \log \left(\frac{z}{1-z}\right)
 \left(2 (\text{Li}_2(z)+\zeta_2)+\log^2(1-z)\right)\,, \\
  g_2^{(3)} &= -12 \left( \text{Li}_3(z)+\text{Li}_3\left(-\frac{z}{1-z}\right) \right)
  + 6 \, \text{Li}_2(z) \log(1-z) + \log^3(1-z) \,,  \\
  g_3^{(3)} &= 6 \log(1-z) \, (\text{Li}_2(z)-\zeta_2)
  - 12 \, \text{Li}_3(z) + \log^3(1-z) \,, \quad
  g_4^{(3)} = \text{Li}_3\left(-\frac{z}{1-z}\right) - 
  3 \, \zeta_2 \log(z) + 8 \, \zeta_3 \,, \\
  g_5^{(3)} & = - 8 \left( \text{Li}_3\left(-\frac{\sqrt{z}}{1-\sqrt{z}}\right)
  + \text{Li}_3\left(\frac{\sqrt{z}}{1+\sqrt{z}}\right) \right)
+ 2 \text{Li}_3\left(-\frac{z}{1-z}\right) + 4 \zeta_2 \log (1-z) +\log \left(\frac{1-z}{z}\right)
     \log^2\left(\frac{1+\sqrt{z}}{1-\sqrt{z}}\right) \,.
\end{align}
}%
The single color coefficients are 

{\footnotesize
\begin{align}
& B_{\text{lc}}(z) =  \frac{122400 z^7-244800 z^6+157060 z^5-31000 z^4+2064 z^3+72305 z^2-143577 z+63298}{1440 (1-z) z^4} \nonumber  \\
& -\frac{-244800 z^9+673200 z^8-667280 z^7+283140 z^6-48122 z^5+2716 z^4-6201 z^3+11309 z^2-9329 z+3007}{720 (1-z) z^5} {g^{(1)}_1} \nonumber  \\
& -\frac{244800 z^8-550800 z^7+422480 z^6-126900 z^5+13052 z^4-336 z^3+17261 z^2-38295 z+19938}{720 (1-z) z^4} {g^{(1)}_2} \nonumber \\
 & +\frac{ 4 z^7+10 z^6-17 z^5+25 z^4-96 z^3+296 z^2-211 z+87}{24 (1-z) z^5} {g^{(2)}_1} \nonumber  \\
& +\frac{-40800 z^8+61200 z^7-28480 z^6+4040 z^5-320 z^4-160 z^3+1126 z^2-4726 z+3323}{120 z^5} {g^{(2)}_2} \nonumber  \\
& -\frac{1-11 z}{48 z^{7/2}} {g^{(2)}_3}
  -\frac{120 z^6+60 z^5+160 z^4-2246 z^3+8812 z^2-10159 z+4193}{120 (1-z) z^5} {g^{(2)}_4} \nonumber \\
& - 2 \left(85 z^4-170 z^3+116 z^2-31 z+3\right) {g^{(3)}_1}
 +\frac{-4 z^3+18 z^2-21 z+5}{6 (1-z) z^5} {g^{(3)}_2}
 +\frac{z^2+1}{12 (1-z)} {g^{(3)}_3},
  \\ \nonumber \\
& B_{\text{nlc}}(z)  = \frac{{57600 z^7-115200 z^6+75748 z^5-17359 z^4+902 z^3+14966 z^2-27552 z+9320}}{720 (1-z) z^4}
\nonumber \\
& -\frac{{-115200 z^9+316800 z^8-321680 z^7+147846 z^6-31035 z^5+3225 z^4-3571 z^3+11322 z^2-12412 z+4880}}{360 (1-z) z^5} {g^{(1)}_1}
\nonumber \\
& -\frac{{230400 z^8-518400 z^7+412960 z^6-138600 z^5+18696 z^4-742 z^3+10971 z^2-25029 z+11424}}{720 (1-z) z^4} {g^{(1)}_2}
\nonumber \\
& +\frac{{-91 z^7+235 z^6-184 z^5+15 z^4-140 z^3+721 z^2-760 z+314}}{120 (1-z) z^5} {g^{(2)}_1}
\nonumber \\
& +\frac{{-19200 z^8+28800 z^7-14680 z^6+2660 z^5-340 z^4-40 z^3+315 z^2-1431 z+952}}{60 z^5} {g^{(2)}_2}
\nonumber \\
& +\frac{{960 z^4-160 z^3+992 z^2+547 z+1435}}{480 z^{7/2}} {g^{(2)}_3}
 -\frac{{-120 z^6+120 z^5-130 z^4-585 z^3+2647 z^2-3143 z+1266}}{60 (1-z) z^5} {g^{(2)}_4}
\nonumber  \\
& +\frac{{640 z^6-1920 z^5+2196 z^4-1196 z^3+318 z^2-42 z+3}}{4 (1-z) z} {g^{(3)}_1}
 +\frac{{2 z^7-3 z^6+3 z^5-z^4-z^3+9 z^2-9 z+1}}{12 (1-z) z^5} {g^{(3)}_2}
\nonumber \\
& -\frac{{(1-2 z) \left(z^2-z+1\right)}}{2 (1-z) z} {g^{(3)}_4}
 -\frac{{2 z^5-z^4+2 z^3+z^2+3}}{4 z^4} { g^{(3)}_5},
  \\ \nonumber \\
& B_{N_f}(z)  =\frac{{7200 z^7-14400 z^6+8852 z^5-1568 z^4+48 z^3+1825 z^2-4115 z+2050}}{144 (1-z) z^4}
 \nonumber  \\
& -\frac{{72000 z^9-198000 z^8+193040 z^7-77700 z^6+10960 z^5-100 z^4-489 z^3+3269 z^2-4801 z+1801}}{360 (1-z) z^5} {g^{(1)}_1}
 \nonumber  \\
& +\frac{{36000 z^8-81000 z^7+60520 z^6-16650 z^5+1190 z^4+10 z^3+428 z^2-939 z+561}}{180 (1-z) z^4} {g^{(1)}_2}
 \nonumber  \\
& +\frac{{-z^7-4 z^3+18 z^2-24 z+9}}{6 (1-z) z^5} {g^{(2)}_1}
 -\frac{{-12000 z^8+18000 z^7-7840 z^6+920 z^5+72 z^2-222 z+187}}{60 z^5} {g^{(2)}_2}
 \nonumber \\
& +\frac{{1-3 z}}{48 z^{7/2}} {g^{(2)}_3}
 +\frac{{ 8 z^3-66 z^2+71 z+7}}{60 (1-z) z^5} {g^{(2)}_4}
 +2 {\left(50 z^4-100 z^3+66 z^2-16 z+1\right)} {g^{(3)}_1}.
\end{align}
}

Let us remark that the unpublished analytic result for the contribution proportional to the number of the light quark flavors $B_{N_f}(z)$ was independently obtained by one of the authors (L.\,D.) and his collaborator Marc Schreiber as early as 2004. The confirmation of this result using our computational framework at an early stage of this work was very encouraging.

We have validated our result against the output of \textsc{Event2} after sampling over $10^9$ points and compared its asymptotic properties to the partial results available in the literature. In the collinear limit ($z\to0$) our  $\log(z)/z$ term agrees with the jet calculus prediction from \cite{Konishi:1978yx,Konishi:1978ax,Richards:1983sr}, while in the back-to-back limit ($z\to1$) terms enhanced by $1/(1-z)$ agree with the results from the NNLL soft gluon resummation \cite{deFlorian:2004mp}. The $\text{N}^3\text{LL}$ soft resummation results should become available in the near future, since the corresponding factorization formula is already known \cite{Moult:2018jzp}.

Last but not least, we also studied the full NLO result in the unphysical limit $z \to \infty$ by analytically continuing the formula initially valid only in the physical region $0\leq z \leq 1$. It is remarkable that each color coefficient is suppressed at least as $1/z^3$, which requires an intricate cancellation between many terms of the full result.

\section{Summary}

The analytic structure of QCD event shape observables beyond LO is still poorly understood. While it is possible to calculate such observables at high precision using numerical codes, fully analytic calculations in the whole range of the given observable require the evaluation of complicated phase-space integrals with involved measurement functions. For many years such calculations were deemed too complicated, but owing to the recent progress in our understanding of IBP reduction, unitarity methods and differential equations, they are finally within reach of modern computers with server-grade CPUs running efficient codes for automatic calculations.

In our recent publication \cite{Dixon:2018qgp}, we employed the publicly available packages \fire, \litered and \fuchsia to obtain the fully analytic result for the EEC in $e^+ e^-$ annihilation at NLO. This result promotes EEC to the very first QCD event shape observable known analytically beyond LO and raises hopes that our method can be used to obtain new analytic results also for more complicated event shape observables, such as $C$-Parameter or Thrust.

\section*{Acknowledgments}

V.\,S. would like to thank the organizers of LL2018 for the opportunity to present 
this work despite the late submission. The work of V.\,S. was supported in part by the National Science Foundation of China (11135006, 11275168, 11422544, 11375151, 11535002) and the Zhejiang University Fundamental Research Funds for the Central Universities (2017QNA3007).
The Feynman diagrams in this proceeding were created with \textsc{JaxoDraw} \cite{Binosi:2003yf}.

\bibliographystyle{JHEP}
\bibliography{inspire}

\end{document}